\documentstyle[emulateapj]{article}

\submitted{Accepted for publication in the Astrophysical Journal, Letters}
\begin{document}

\title{LRG J0239-0134: A RING GALAXY OR A PAIR OF SUPERBUBBLES AT z=1 ?}

\author{Yoshiaki Taniguchi$^{1, 2}$ and Takashi Murayama$^1$}

\affil{$^1$ Astronomical Institute, Graduate School of Science, 
       Tohoku University, Aramaki, Aoba, Sendai 980-8578, Japan \\
       $^2$ Institute for Astronomy, University of Hawaii, 2680
       Woodlawn Drive, Honolulu, HI 96822}

\begin{abstract}
The unusual morphology of LRG J0239$-$0134 at $z$ =1.062 has been interpreted
as a ring galaxy. 
Here we propose an alternative idea that the ring-like
morphology is attributed to a pair of superbubbles driven by the intense
starburst in the central region of this galaxy.
Supporting evidence for a superbubble model is;
1) the poststarburst nature in the central body suggests that
a burst of supernova explosions could occur at least $\sim 10^7$ yr ago,
2) the dark lane seen in the central body suggests that we observe this
object from a nearly edge-on view, and, 3) the ring-like morphology
is not inconsistent with an idea that it is a pair of superbubbles.
All these pieces of evidence for the superbubble model seem circumstantial.
However, if this is the case, this galaxy provides an important example
of the superwind activity at high redshift.
\end{abstract} 


\section{INTRODUCTION}

The greatest observational capability of the
{\it Hubble Space Telescope (HST)} has been providing us a lot of invaluable
information on the evolution of galaxies (e.g., Williams et al. 1996;
Lanzetta, Fern\'andez-Soto, \& Yahil 1996; van den Bergh et al. 1996;
Cowie, Hu, \& Songaila 1995; van Dokkum et al. 1999).  
In addition, massive and clusters of galaxies have been also used as
natural gravitational telescopes to investigate properties of high-$z$
galaxies [Smail et al. 1996; Colley, Tyson, \& Turner 1996, Franx et al. 1997; 
Soucail et al. 1999 (hereafter S99); P\'ello et al. 1999; 
B\'ezecourt et al. 1999].

Since optical observations of high-$z$ galaxies probe their
redshifted ultraviolet radiation, we must be careful in interpreting the data
because active star-forming regions may be selectively shown up, making often
galaxy morphology knotty or irregular (Bohlin et al. 1990; Giavalisco et al.
1996; Hibbard \& Vacca 1997). 
However, on one hand, it seems likely that young galaxies
have more chaotic morphological properties because they are dynamically younger
than present-day galaxies.
For example, galaxies with chain-like morphology have been
found in {\it HST} deep images (Cowie et al. 1995; van den Bergh et al.
1996; Abraham 1997; Bunker et al. 2000; Taniguchi \& Shioya 2001).
It is also mentioned that 
more frequent collisions between/among galaxies (or subgalactic
building blocks) result in peculiar morphology of high-$z$ galaxies
(e.g., van den Bergh et al. 1996).

Among galaxies with peculiar morphologies, one interesting object is
a ring-like galaxy at $z = 1.062$ (LRG J0239$-$0134) found as an optical
counterpart of the 850 $\mu$m source, SMM J02399$-$0134
(Smail et al. 1998: S99). 
Such ring galaxies are not so rare in the local universe and their
formation mechanism is now well known; the axial penetration of a
smaller companion galaxy to a disk galaxy
(Theys \& Spiegel 1976, 1977; Lynds \& Toomre 1976;
Appleton \& Marston 1997 and references therein).
Therefore, if LRG J0239$-$0134 is a ring galaxy, it would have a less-luminous 
companion galaxy. Although two galaxies are close to LRG J0239$-$0134
(see Figure 1),
one of them, \#32 in Mellier et al. (1998), belongs to the cluster
of galaxies Abell 370 ($z = 0.37$) and the other galaxy A in S99,
or L4 in Barger et al. (1999), has a redshift of 0.42.
Thus, these two galaxies have no relation to  LRG J0239$-$0134.
Although another two very faint objects (B and C in Figure 1) are also
seen to the south of LRG J0239$-$0134,
they appear too faint to be a colliding partner.
The most probable candidate for the colliding partner seems 
an object seen just to the north of the main  body of LRG J0239$-$0134
(D in Figure 1) although we cannot rule out the possibility that 
this object might not be separate from the main body of LRG J0239-1034.
However, there is no spectroscopic confirmation.
Hence it seems still uncertain whether or not LRG J0239$-$0134 is
a {\it usual} ring galaxy made by the axial penetration of a companion
galaxy.
In this Letter, we propose an alternative idea that the ring-like
morphology is attributed to a pair of superbubbles driven by the intense
starburst in the central region of this galaxy.

\section{SUPERBUBBLE MODEL FOR LRG J0239$-$0134}

\subsection{Observational Properties of LRG J0239$-$0134}

S99 reported the detailed observational properties
of LRG J0239$-$0134. Since this galaxy is magnified and distorted by
the gravitational shear induced by the cluster Abell 370 and the 
nearby elliptical galaxy \#32, they made source reconstruction of 
LRG J0239$-$0134. In their model, the magnification factor is derived
as $2.5 \pm 0.2$ but it varies from 2.1 to 3.6 depending on the
distance to the galaxy \#32. Following S99, we adopt a Hubble constant
$H_0 = 50 h_{50}$ km s$^{-1}$ Mpc$^{-1}$, a deceleration $q_0 = 0.5$,
and a cosmological constant $\Lambda = 0$.

The intrinsic $B$-band absolute luminosity is $L_{\rm B} = 1.3 \times 
10^{12} h_{50}^{-2} L_{\rm B, \odot} \simeq 6.0 \times 10^{11} 
h_{50}^{-2} L_\odot$.
The ring-like feature has an intrinsic 
radius of $r_{\rm ring} = 7.7 \pm 0.4 h_{50}^{-1}$ kpc. 
Its major-to-minor axis ratio is 
$a/b = 0.76$. If the ring is a circular one, the axial ratio suggests 
that the inclination angle is $i = 40^\circ$.
The mid-infrared (MIR) luminosity at $\lambda_{\rm rest} = 3.3 \mu$m is
$L_{\rm 3.3 ~\mu m} \simeq 4 \times 10^{10} L_\odot$, being 
ten times as luminous as that of the Cartwheel galaxy which is one of
well-known ring galaxies in the local universe
(Charmandaris et al. 1999).

S99 made optical spectroscopy and detected two emission lines,
[O {\sc ii}]$\lambda$3727 and [Ne {\sc v}]$\lambda$3426.
Another emission feature is marginally identified as 
[Ne {\sc iii}]$\lambda$3869. 
These emission-line properties suggest that LRG J0239$-$0134 has an
active galactic nucleus.
Recently, X ray (0.3 -- 7 keV) emission has been detected by {\it Chandra} from 
this galaxy (Bautz et al. 2000). 
Bautz et al. (2000) estimated the unobscured X-ray luminosity,
$L_{\rm X} \sim 5 \times 10^{11} h_{50}^{-2} L_\odot$ and found that
the X-ray to submm flux ratio is similar to that of NGC 6240 which is
one of luminous infrared galaxies harboring an active galactic nucleus
(e.g., Iwasawa 1999 and references therein).

Stellar continuum emission is dominated by the central body of
LRG J0239$-$0134. Since Balmer lines  (H$\delta$, H$\epsilon$, H8,
and H9) and Mg {\sc ii} are observed as absorption, it is suggested that
A and/or B type stars are
dominant stellar sources in the central body. They are often
interpreted as evidence for poststarburst (e.g., Taniguchi et al. 1996;
Brotherton et al. 1999; Taniguchi, Shioya, \& Murayama 2000).
S99 noted that [O {\sc ii}]$\lambda$3727 emission is spatially extended,
suggesting that a part of the ring-like feature is also an emission-line
region. There is no significant velocity difference between
the extended emission and the central body emission
($\Delta v \lesssim 700$ km s$^{-1}$).
Since [Ne {\sc v}]$\lambda$3426 does not show such an extended feature,
the extended [O {\sc ii}] emission-line region is perhaps photoionized 
by newly formed massive stars in the ring.
Finally we mention that there is an evident dark lane along the
major axis of the central body; however, we would like to note that 
the dark-lane hypothesis may be weakened if object D belongs to the
main body of LRG J0239$-$0134. 

\subsection{Superbubble Model for LRG J0239$-$0134}

First we give a summary of supporting evidence for a superbubble model.
1) The poststarburst nature in the central body suggests that 
a burst of supernova explosions could occur at least $\sim 10^7$ yr ago.
2) The dark lane seen in the central body suggests that we observe this
object from a nearly edge-on view. Note that we would see this object 
from a nearly face-on view if the ring-like feature is a really ring.
And, 3) the ring-like morphology in the raw image of S99
is not inconsistent with an idea that it is a pair of superbubbles.
Indeed, the figure 8-shaped morphology in the raw image of S99 looks
quite similar to that of Arp 220 (see Heckman, Armus, \& Miley 1987).

Now we apply the superbubble model
(McCray \& Snow 1979; Koo \& McKee 1992a, 1992b;
Heckman et al. 1996; Shull 1995; see also for a review on
observational properties of nearby superwinds, Heckman, Armus, \&
Miley 1990) to the case of LRG J0239$-$0134.
Here it is noted that we ignore the effect of gravity for simplicity.
The radius of the superbubble (i.e., shocked shells) at time $t$
(in units of $10^7$ yr) is then

\begin{equation}
r_{\rm shell}  \sim 2.8
L_{\rm mech, 43}^{1/5}
n_{\rm H, 0}^{-1/5}
t_{7}^{3/5} ~~ {\rm kpc},
\end{equation}
where $L_{\rm mech}$ is the mechanical luminosity
released collectively from the supernovae in the central starburst
in units of $10^{43}$ ergs s$^{-1}$ and $n_{\rm H}$ is the average
hydrogen number density of the
ISM, assumed constant in units of 1 cm$^{-3}$;
this average density is obtained by assuming that the hydrogen gas
with a mass of $10^{11} M_\odot$ is distributed uniformly in 
a sphere with a radius of 10 kpc. 
The inferred infrared luminosity of LRG J0239$-$0134
[a few times $10^{12} L_\odot$ under $H_0 = 75$ km s$^{-1}$ Mpc$^{-1}$ and 
$q_0 = 0$ (Sanders \& Mirabel 1996): see section 3.1]
is nearly comparable to those of ultraluminous infrared galaxies (ULIGs)
such as Arp 220. For Arp 220, one of the well-known superwind galaxies,
Heckman et al. (1996) derived 
$L_{\rm mech} \simeq 10^{43}$ ergs s$^{-1}$. Therefore, we have adopted
this value as a fiducial one\footnote{
For a forming elliptical galaxy with a stellar mass 
$M_{\rm stars} = 10^{11} M_\odot$,
radius  $r \simeq $ 10 kpc and ${n}_{\rm H} \sim 1$ cm$^{-3}$
(see Saito 1979; Arimoto \& Yoshii 1987), we expect
$N_{\rm SN} \sim 3 \times 10^9$ stars that explode as supernovae.
Since most of these massive stars might be formed during the first 0.5 Gyr
(= $t_{\rm GW}$), we obtain 
$L_{\rm mech} \sim \eta ~ E_{\rm SN} ~ N_{\rm SN} / t_{\rm GW} \sim 10^{43} ~
{\rm erg~ s}^{-1}$ where $E_{\rm SN}$ is the total energy of a single supernova
($10^{51}$ erg) and $\eta$ is the efficiency of the kinetic energy deposited
to the ambient gas ($\sim$ 0.1; Dyson \& Williams 1980).}.
It is also noted that the luminosity dependence is fairly weak; i.e.,
$r_{\rm shell} \propto L_{\rm mech}^{1/5}$.

Using the observed ring radius, $r_{\rm ring} = 7.7 \pm 0.4 h_{50}^{-1}$ kpc,
we estimate the age of the superbubble is $5.4 \times 10^7$ yr.
Since the elapsed time from the onset of the starburst exceeds
$\sim 1 \times 10^7$ yr, Balmer absorption dominates over Balmer
emission and thus the Balmer lines are observed as absorption lines
(e.g., Taniguchi et al. 2000). This appears consistent with the
observation.

The expansion velocity of the superbubble at time $t$
(in units of $10^7$ yr) is

\begin{equation}
v_{\rm shell} \sim 164 
L_{\rm mech, 43}^{1/5}
n_{\rm H, 0}^{-1/5}
t_{7}^{-2/5} {\rm km ~ s}^{-1}.
\end{equation}
Using $t_7 = 5.4$, we obtain $v_{\rm shell} \sim 84$ km s$^{-1}$.
Since we assume that the ring-like feature is a pair of superbubbles,
we see mostly the tangential section of the superbubbles.  
Therefore, their velocity may be nearly the same as that of the systemic
velocity of the central body. Even if we observe some expanding blobs
along the line of sight, the above shock velocity seems too slow to be
detected in low-resolution spectroscopy.

The superbubble model discussed here treats the
spherical symmetric case for simplicity.
However, actual galaxies may have oblate spheroidal isopotential surfaces
to some extent even when they are in a forming phase.
In such a case, the superbubble will blow out along the minor
axis of the galaxy (Tomisaka \& Ikeuchi 1988; De Young \& Heckman 1994).
Therefore, it is likely that a pair of superbubbles can be formed
toward the two polar directions. This is schematically illustrated in Figure 2.

In the context of the superbubble model, the ring-like feature 
could originally consist of gas clouds. However, as described in 
Taniguchi, Trentham, \& Ikeuchi (1999), star clusters may form
as pressure-confined condensations in a shell (or a bubble).
Therefore, the ring-like
feature can be seen in the optical continuum emission as well as 
in the line emission arising from nebulae photoionized by newly
formed massive stars.
Star formation in young galaxies (e.g., an initial starburst)
is expected to be much more intense than
typical nuclear starbursts in the local universe.
Further, the central starburst region in young galaxies
may still be surrounded by a lot of gas.
Hence, any effects of superbubbles
must be much more serious in such young galaxies than in 
galaxies in the local universe.

\section{DISCUSSION}

\subsection{LRG J0239$-$0134 as an Ultraluminous Infrared Galaxy}

Photometric data from rest-frame ultraviolet (UV) to MIR
of LRG J0239$-$0134 were obtained by S99. In addition to these data, 
Barger et al. (1999) reported its 850 $\mu$m flux, 11 mJy.
Using these data, we investigate the overall spectral energy distribution 
(SED) of LRG J0239$-$0134. The SED is shown in Figure 3 together with 
those of NGC 6240 (Griffith et al. 1995; Lisenfeld et al. 1996, 2000; 
Klaas et al. 1997; Spinoglio et al. 1995; White \& Becker 1992)
and  Arp 220 (Rigopoulou, Lawrence, \& Rowan-Robinson 1996;
Klaas et al. 1997).
As shown in Figure 3, the SED of LRG J0239$-$0134 is much similar to 
that of NGC 6240 than to that of Arp 220. This makes sense because
NGC 6240 harbors a luminous active galactic nucleus while Arp 220
may not. Fitting the two SEDs of NGC 6240 and Arp 220 to the observed
SED of LRG J0239$-$0134,
we obtain the total IR luminosity between 8 $\mu$m and
1000 $\mu$m, $L$(8-1000$\mu$m) $\sim
 7.9 \times 10^{12} h_{50}^{-2} L_\odot$ and 
 $8.6 \times 10^{12} h_{50}^{-2} L_\odot$, respectively;
in the case of $ h_{50}=1.5$, $L$(8-1000$\mu$m) 
$\sim 3.5 \times 10^{12} L_\odot$ and 
$\sim 3.8 \times 10^{12} L_\odot$, respectively. 

For nearby starburst galaxies including some local ULIGs, Mouri et al. (1990)
found a relation of $L_{\rm FIR} \sim 10^3 L_{\rm 3.3\mu m}$
where $L_{\rm FIR}$ is the FIR luminosity derived from rest-frame
60 $\mu$m and 100 $\mu$m fluxes (e.g., Sanders et al. 1988;
Sanders \& Mirabel 1996).
Using the observed $L_{\rm 3.3\mu m}$ (S99), we would obtain
$L_{\rm FIR} \sim 4 \times 10^{13} L_\odot$.
However, from the SED fitting, we obtain
$L_{\rm FIR} \sim 8 \times 10^{12} h_{50}^{-2} L_\odot$ at most,
being lower by a factor of five  than the above expected value.
Although the SED concordance between high-$z$ submm sources and local
ULIGs has been claimed (Ivison et al. 1998; see also Trentham 2000),
it is suggested that LRG J0239$-$0134 is a different population from
local {\it starburst-dominated} 
ULIGs though its IR luminosity is comparable to those of
the ULIGs\footnote{
Typical ring galaxies in the local universe have FIR luminosities of
$L_{\rm FIR} \sim 4 \times 10^{10} h_{50}^{-2} L_\odot$
(Appleton \& Struck-Marcell 1987), being much fainter than that of
LRG J0239$-$0134. On the other hand, they have ring radii of
20 - 40 kpc (Appleton \& Struck-Marcell 1987), being significantly
larger than that of LRG J0239$-$0134.}.

\subsection{Concluding Remarks}

We have presented our superbubble model for the unusual ring-like
galaxy LRG J0239$-$0134 at $z = 1.062$. All pieces of supporting evidence
for the superbubble model seem circumstantial. However, 
if this is the case, this galaxy provides an important example
of the superwind activity at high redshift.

Similar examples may be  the
extended Ly$\alpha$ blobs (LABs) at $z \approx 3.1$ found by
Steidel et al. (2000) because their observational properties are also
interpreted in terms of superwind activities (Taniguchi \& Shioya 2000;
cf. Haiman, Spaans, \& Quataert 2000; Fardal et al. 2000).
Taniguchi \& Shioya (2000) proposed an evolutionary connection from
dust-enshrouded (or dusty) submm sources (hereafter DSSs; e.g.,
Hughes et al. 1998; Barger et al. 1998; Barger et al. 1999; Ivison et al. 
2000 and references therein) to the LABs.
It is interesting to note that LRG J0239$-$0134 can be considered
as a link between the DSSs and the LABs.
UV imaging of LRG J0239$-$0134 will be important to probe
its extended Ly$\alpha$ emission.

Our model predicts that the central body of LRG J0239$-$0134 is rotating
along the minor axis. Further, the ring-like regions are affected by
shock heating to some extent. High-resolution optical 
spectroscopy will be useful in examining these predictions.

\vspace {0.5cm}

We would like to thank Kazushi Iwasawa and Yasuhiro Shioya for
useful discussion and comments.
We would also like to thank an anonymous referee for useful comments
and suggestions.
YT thanks Rolf-Peter Kudritzki, Bob McLaren, and Dave Sanders
at Institute for Astronomy, University of Hawaii
for their warm hospitality. 
This work was financially supported in part by
the Ministry of Education, Science, and Culture, Japan
(grants 10044052, and 10304013).



\begin{figure}
\epsfysize=12cm \epsfbox{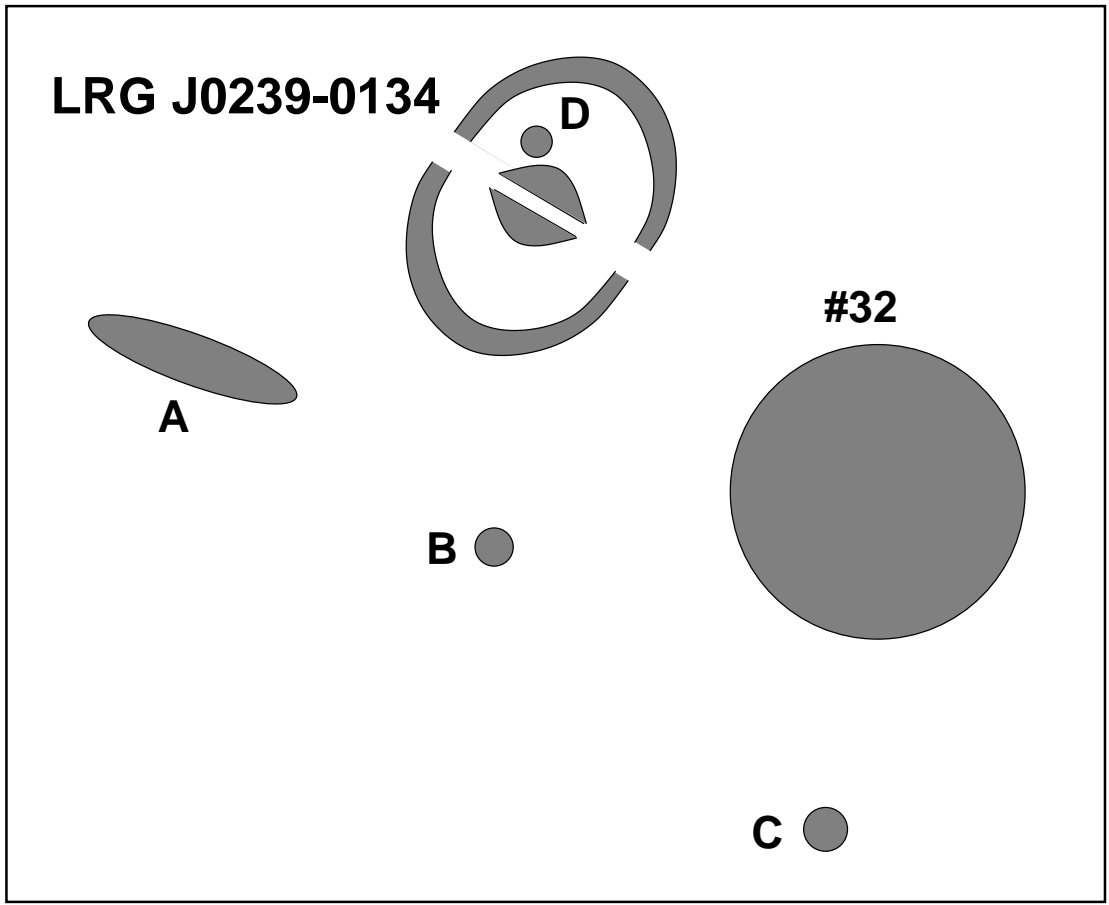}
\caption[]{
A schematic illustration of the environ of LRG J0239$-$0134
based on the raw image given in Figure 1 of S99.
This Figure is made for readers' convenience to see the
relative positions of objects B, C, and D, and thus
does not scale accurately.
See for more details Figures 1 and 4 in S99.
\label{fig1}
}
\end{figure}

\begin{figure}
\epsfysize=8cm \epsfbox{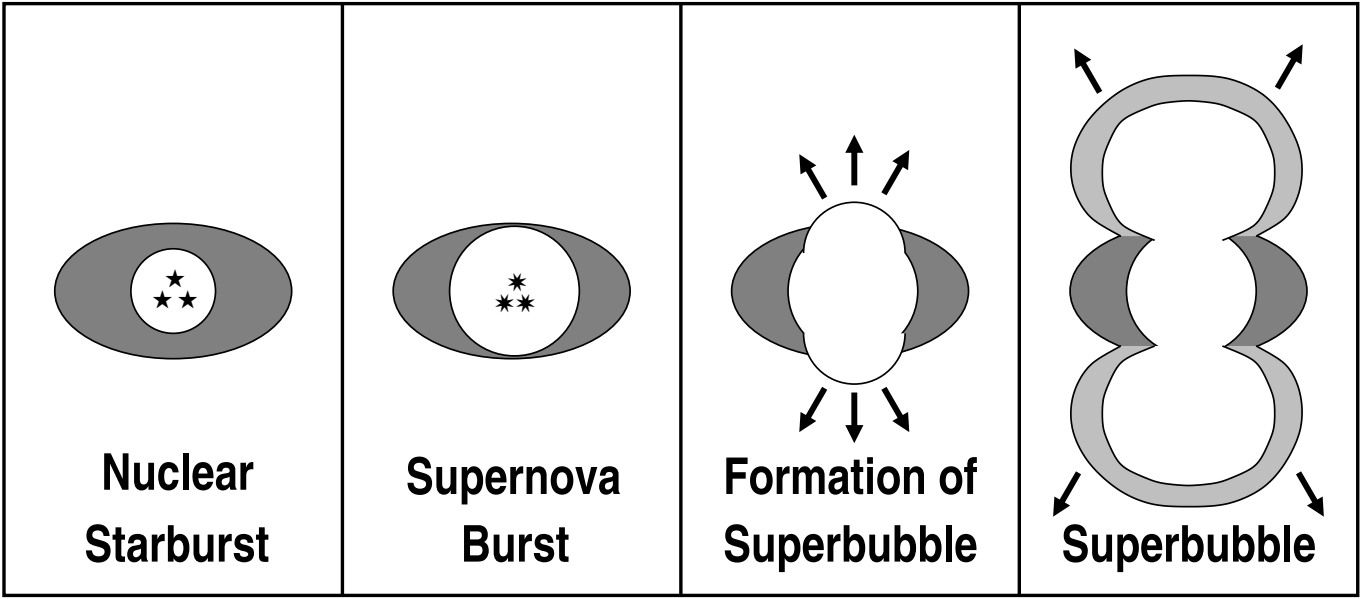}
\caption[]{
A schematic illustration of our superbubble model
for LRG J0239$-$0134 (see also De Young \& Heckman 1994).
\label{fig2}
}
\end{figure}

\begin{figure}
\epsfysize=10cm \epsfbox{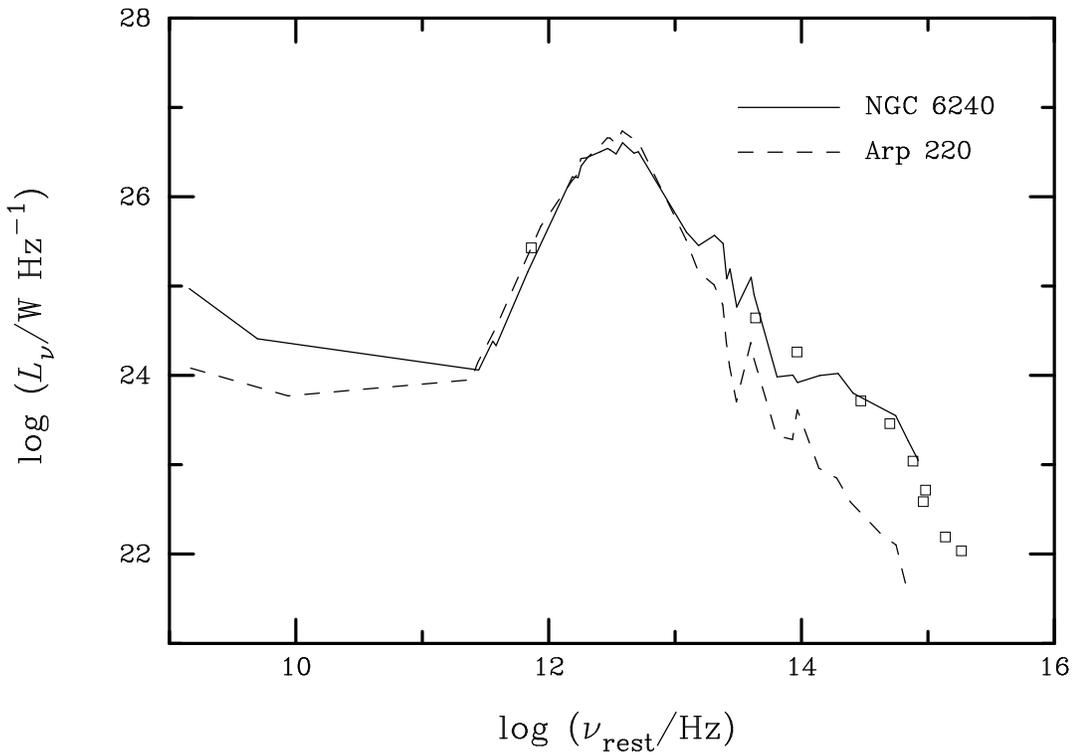}
\caption[]{
Spectral energy distribution of LRG J0239$-$0134. The photometric data of
LRG J0239$-$0134 are shown by open boxes.  For comparison,
we show the SEDs of NGC 6240 (solid line) and Arp 220 (dashed line).
\label{fig3}
}
\end{figure}

\end{document}